\documentstyle[epsf,wrapfig]{aipproc}

\begin{document}

\newcommand{\mr}{\mathrm}
\newcommand{\fluxunit}{\mathrm{cm}^{-2}\mathrm{s}^{-1}}

\title{Recent Observations of $\gamma$-rays 
above 1.5 TeV from Mkn 501 with the 
HEGRA 5 m$^2$ Air \v{C}erenkov Telescope}

\author{D. Kranich$^1$, T. Deckers$^2$, E. Lorenz$^1$, D. Petry$^1$,\\ G. Rauterberg$^2$ and the HEGRA Collaboration}
\address{$^1$ Max-Planck-Institut f\"ur Physik, F\"ohringer Ring 6,  80805 M\"unchen, Germany\\
$^2$ Institut f\"ur Kernphysik, Olshauserstr. 40, 24118 Kiel, Germany}

\maketitle

\begin{abstract}
Since February 1997 the BL Lac object Mkn 
501 is in a ``high state'' of $\gamma$-ray emission. The HEGRA collaboration has 
studied Mkn 501 with their air \v{C}erenkov telescopes on La Palma. Here 
we report on observations with the 5 m$^2$ telescope (threshold $\approx$ 1.5 TeV) 
operated in a stand alone mode. We observed a rapidly varying flux 
between 0.5 to 6 times of that from the Crab Nebula. On average a Mkn  
501 flux of $(2 + 1.3 - 0.5) \times 10^{-11} \fluxunit$ has been determined. The spectrum 
extends at least up to 10 TeV with an integral power law coefficient of 
$1.8 \pm 0.2$ and seems to be steeper than in 1996.
\end{abstract}

\section*{Introduction}
The AGN Mkn 501 is known to be a VHE $\gamma$-ray emitter \cite{quinn1,bradbury2} while it has 
not been observed by EGRET in the HE domain above 100 MeV. It is a variable 
source with a significant increase in intensity since its discovery by \cite{quinn1} in 1995. 
Since about February 1997 the source is in a 'high state' of emission showing 
strong variability \cite{breslin3}. The HEGRA collaboration has studied Mkn 501 since early 
March with their air \v{C}erenkov telescopes (ACT). The ACTs are part of a detector 
complex for the study of cosmic rays  over an energy range between $\approx$ 500 GeV 
and $10^{17}$ eV. The installation is located on the Roque de los Muchachos, La 
Palma, Islas Canarias (28.8$^\circ$ N, 17.8$^\circ$ W, 2200 m a.s.l.). In total, 6 ACTs (5 of 8.5 
m$^2$ and one with 5 m$^2$ mirror area) are in operation. Four of the 8.5 m$^2$ 
ACTs were operated in the so-called stereo-mode while the 5 m$^2$ ACT 
(hereafter called CT1) was operated in a stand-alone mode. Here we report 
on observations with CT1 while the results from the stereo system are reported 
elsewhere at this symposium and first data are already submitted for publication 
\cite{aharonian4}. Compared to the system, CT1 has a higher threshold (1.5 vs. $\approx$ 0.5 TeV) and 
only a 3.25$^\circ$ camera of 127 pixels, but was available for longer observation time 
starting at larger zenith angles. Also, the DAQ has a larger dynamic range for 
pulse height recording and observations can be carried out under the presence of 
moon light \cite{raubenheimer5}. Technical details of CT1 are given in \cite{mirzoyan6}.

\section*{Data sample and analysis}
Since the 9th of March (MJD 50517) the source Mkn 501 has been 
observed continuously whenever weather conditions permitted it ($\approx$ 30 \% of time 
lost). During periods of moonlight, observations were carried out with reduced 
photomultiplier (PM) HV resulting in a 30-60 \% higher threshold. Only during the days of full 
moon and the source being closer than 30$^\circ$ to the moon the measurements had to 
be stopped because PM anode currents exceeded safe limits. 
Observation times were typically 2-5 hours per night with raw-data  trigger rates of 0.5 - 
0.9 Hz (zenith angle dependent). The trigger condition was a coincidence of 2 
pixels (of 127) exceeding each 17 photoelectrons (PE). After MJD 50568 the threshold was lowered
to 13 PE by increasing the HV by 6 \% resulting in a raw- data trigger rate of up to 1.8 Hz.

\begin{figure}[b]\centering \leavevmode
        \epsfxsize=14.5cm
        \epsffile{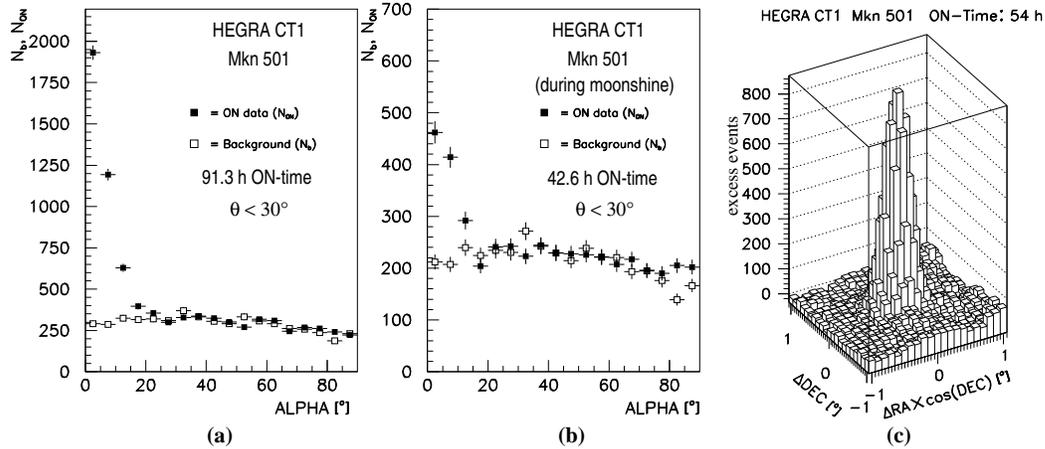}
        \caption{\label{fig1}
               {\bf (a)} The ALPHA distribution for on-source data taken during moonless night and background after
                 all cuts except that on ALPHA. The angle ALPHA describes the orientation of the
                shower image and is expected to be small for $\gamma$-showers
                from the direction of the assumed source-position.
                 At ALPHA $< 10^\circ$ there are 2554 excess events (40.4 $\sigma$). 
                {\bf (b)} like (a) but for the data taken at times when the moon was above the horizon. 
                 At ALPHA $< 10^\circ$ there are 455 excess events (12.4 $\sigma$). 
                {\bf (c)} Excess event distribution obtained by varying the assumed source position by up to $\pm 1^\circ$
                along the RA and DEC axes (see text).
                }
\end{figure} 

 The data were recorded entirely in the so-called tracking mode. The filtering and 
processing  of the data and background determination followed the prescription given in \cite{bradbury2}. After applying 
some filter cuts to remove accidental triggers etc., standard Hillas image 
parameters were calculated. For the selection of $\gamma$-ray candidates modified Supercuts, so-called dynamical
supercuts \cite{kranich8} were applied. These cuts vary with energy 
(equivalent parameter: SIZE), shower impact parameter (equivalent parameter: 
DIST) and the zenith angle $\theta$.
 The cuts were at first optimised with Monte Carlo 
(MC) data and then fine-tuned on previously recorded data from the Crab Nebula.

 Fig. \ref{fig1}a and b show the ALPHA distributions for the total data with $\theta < 30^\circ$
which was used to produce the light curve shown in figure \ref{fig3} 
together with the background determined from previously recorded off-source data.
Fig. \ref{fig1}c shows a ``lego plot'' of excess events versus assumed source position.
The excess is centered within 
0.02$^\circ$ (= tracking error of CT1) of the nominal Mkn 501 position and has an 
angular spread of $\sigma = (0.10 \pm 0.02)^\circ$. For a short period, CT1 was pointed 0.3$^\circ$ away 
from the source and the data showed the excess with the same shift and identical 
spread. The significance of the daily signal varied between 1 and 15 $\sigma$
depending on the flux and threshold while the excess rate varied 
between 2 and 60 h$^{-1}$. The background, however, remained stable (see fig.  \ref{fig2}).

\section*{Results}
\begin{figure}\centering \leavevmode
        \epsfxsize=14cm
        \epsffile{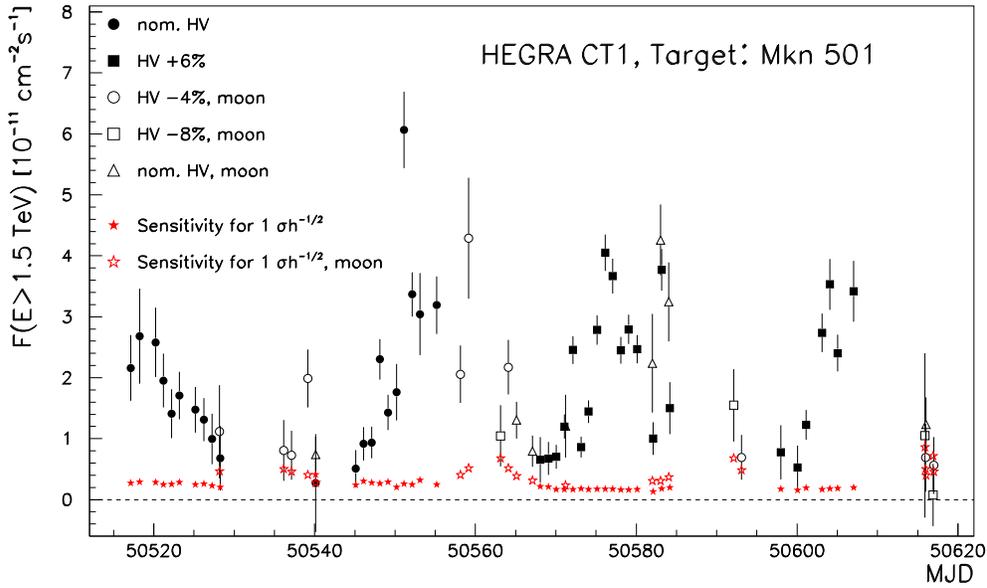}
        \caption{\label{fig2}
                  The preliminary daily flux measurements from March to June 1997 from the data at zenith angle
                  $\theta < 30^\circ$.
                  The different symbols indicate the different data taking conditions. Also indicated is the
                  sensitivity, i.e. the flux corresponding to a $\gamma$-rate which is equal to
                  the squareroot of the background rate in h$^{-1}$; this gives a measure
                  of the stability of the general observing conditions.
                }
\end{figure}

The rates were converted into a flux taking into account the collection area 
(determined by Monte Carlo simulations), change of threshold due to $\theta$, increase of 
threshold due to the different HV settings and the dynamical supercuts (loss of $\gamma$-ray events of $\approx$
40 \%). Details will be given in a forthcoming paper. 
In order to assess the variability of the source, we present here the lightcurve for the data 
at $\theta < 30^\circ$ with an average threshold around 1.7 TeV (Fig. \ref{fig2}). The threshold rises with zenith 
angle and with reduced HV therefore the flux data from the concerned observations were extrapolated to 1.5 
TeV by using a power law coefficient of -1.8 (see below). 
The complete analysis of the data up to $\theta = 60^\circ$ will be 
presented elsewhere. The daily flux showed large variations sometimes 
exceeding the CRAB flux by up to a factor 6. 
We determine
an average flux (MJD 50517 to MJD 50608) of
$$
         F_{\mr{Mkn501, March-June 1997}}(E > 1.5 \mr{TeV}) = (2 +1.3 -0.5) \times 10^{-11} \fluxunit
$$
where the errors given are mainly systematic.       
The comparable flux in 1996 was \cite{bradbury2}
$
F_{\mr{Mkn501, March-August 1996}}(E > 1.5 \mr{TeV}) = (2.3 (\pm 0.4)_{\mr{stat}}(+1.5-0.6)_{\mr{syst}}) \times 10^{-12} \fluxunit
$
and the Crab Nebula flux  \cite{bradbury2} 
$F_{\mr{Crab,1996}}(E > 1.5 \mr{TeV}) = (7.7 (\pm 1.0)_{\mr{stat}}(+4.6-1.9)_{\mr{syst}}) \times 10^{-12} \fluxunit$.

\begin{figure}
\begin{minipage}[t]{8cm}
\epsfxsize=8cm
\epsffile{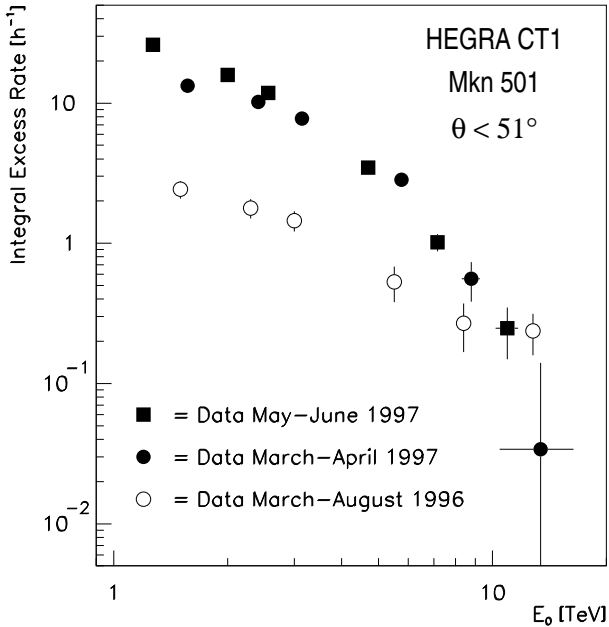}
\end{minipage}
\hfill
\begin{minipage}[b]{4.5cm}
\caption{\label{fig3}
        Integral $\gamma$-candidate excess rates vs. energy threshold $E_0$ for all data taken in 1996 (220 h on-time)
        and 1997 (54 h on-time in the March-April dataset and 80 h in the May-June dataset) at zenith angles $\theta < 51^\circ$.
        The bars are the statistical errors. The threshold was raised by applying a zenith angle dependent
        cut on SIZE. The zenith angle distributions of the three datasets up to $51^\circ$ are similar.
        }
\end{minipage}
\end{figure}

From MJD 50548 to 50554 (April 9 - 15, 1997) EGRET was pointed to the direction of Mkn 501 but 
no evidence for $\gamma$-ray emission above 100 MeV has been found from the 
preliminary analysis \cite{kanbach10} with an upper limit in the range of the previous one of 
$1.7 \times 10^{-7} \fluxunit$ while we measured during the same time a mean flux of 
$ (3 +1.9 -0.8) \times 10^{-11} \fluxunit$. This indicates that the spectrum must be very hard with an 
integral power law index between 100 MeV and 1.5 TeV $< 1$.
In Fig. \ref{fig3} we show the average integral $\gamma$-excess rates as a function of energy thresholds $E_0$  
for the first and the second half of the current observation period
and for the data from 1996 (the 1996 data published in [2] were
reanalysed using the dynamical supercuts which resulted in a 6.8 sigma signal with 494 excess events).
These rates can be turned into fluxes by dividing by the appropriate expected rate which has to be determined
from Monte Carlo studies. Preliminary investigations show that up 
to about 10 TeV the 1997 spectrum can be described by a power law with an integral 
coefficient of $1.8 \pm 0.2$  which is within the errors compatible with the slope of $1.58 \pm 0.51$  of 
the 1996 data but the convergence of the rates at larger energies seems to indicate that the
spectrum in 1997 is steeper than it was in 1996 and that the flux at energies above 10 TeV is essentially unchanged.
This needs further investigations. Both the 1996 and 1997 spectra are incompatible with a large 
cosmological IR background calculated by \cite{salamon11} (and references therein) although a single source 
observation is not sufficient to rule out all possible scenarios.

As Mkn 501 was also observed 
by the Whipple and CAT 
groups (see their contributions to this symposium) and both locations are 
at different longitude it is 
expected that soon nearly 
continuous data will be available 
covering a long period. 
The ongoing observations of Mkn 501
at energies from radio waves to $\approx$ 10 TeV
will provide a unique opportunity to study the multi-wavelength
behaviour of BL Lac objects.
\vspace{-0.2cm}

\subsubsection*{Acknowledgements}
\vspace{-0.2cm}
We thank T. C. Weekes for the early information on the increase in activity 
of Mkn 501. The support of the German BMBF and Spanish CICYT is gratefully 
acknowledged. Also we thank the Instituto de Astrofisica de Canarias for the use 
of the site and the provision of excellent working conditions.

\end{document}